\documentclass[jcp,twocolumn,superscriptaddress,floatfix]{revtex4-1}
\usepackage{epsfig}

\begin{document}
\title{Vapor condensation onto a non-volatile liquid drop }

\author{Levent Inci}
\affiliation{Department of Chemistry, University of Saskatchewan, Saskatoon, SK, S7N 5C9, Canada}

\author{Richard K. Bowles}
\email{richard.bowles@usask.ca}
\affiliation{Department of Chemistry, University of Saskatchewan, Saskatoon, SK, S7N 5C9, Canada}

\date{\today}

\begin{abstract}
Molecular dynamics simulations of miscible and partially miscible binary Lennard--Jones mixtures are used to study the dynamics and thermodynamics of
vapor condensation onto a non-volatile liquid drop in the canonical ensemble. When the system volume is large, the driving force for condensation is low and only a submonolayer of the solvent is adsorbed onto the liquid drop. A small degree of mixing of the solvent phase into the core of the particles occurs for the miscible system. At smaller volumes complete film formation is observed and the dynamics of film growth are dominated by cluster-cluster coalescence. Mixing into the core of the droplet is also observed for partially miscible systems below an onset volume suggesting, the presence of a solubility transition. We also develop a non-volatile liquid drop model, based on the capillarity approximations, that exhibits a solubility transition between small and large drops for partially miscible mixtures and has a hysteresis loop similar to the one observed in the deliquescence of small soluble salt particles. The properties of the model are compared to our simulation results and the model is used to study the formulation of classical nucleation theory for systems with low free energy barriers.
\end{abstract}

\maketitle

\section{Introduction}

Atmospheric aerosols ranging in size from a few molecules to a 100$\mu$m and containing complex mixtures of soluble, insoluble, miscible and immiscible species from a variety of anthropological and natural sources, play a critical role in the microphysics of clouds. Soluble and insoluble particles can act as cloud condensation nuclei (CCN) by providing heterogeneous nucleation sites that lower the free energy barrier for droplet formation and ice nucleation. They also serve as reactants and catalysts in atmospheric chemical cycles. However, the ability of aerosol particles to initiate water uptake is dependent on both the chemical composition of the particle and its size.~\cite{Martin:2000do} Consequently, understanding how composition, the distribution of material inside a particle and particle size impacts its CCN affinity remain important questions.

K\"{o}hler theory of activation~\cite{kolher1,kolher2,kolher3} is based on the stability of the solutions to the Kelvin relation for a volatile solvent vapor in contact with a non-volatile drop and argues condensation begins at the vapor pressure where the drop becomes unstable with respect to particle growth. While nucleation theory provides a more detailed picture of condensation as  a barrier crossing process, the difference in the predictions of the two approaches only becomes apparent for very small, non-volatile drops~\cite{Mirabel:2000cc} and the simplicity of the activation theory is appealing in most practical cases. When the liquids in the drop are ideal, the Kelvin relation has a stable and unstable region in the coexisitence curve but Reiss and Koper\cite{Reiss:1995p7018} found that the Kelvin relation involving a non-ideal liquid solution exhibits an additional unstable branch that is related to the mixing of the two components.  Tanlanquer and Oxtoby\cite{Talanquer:2003p469} used classical density functional theory (DFT) to explore the properties of miscible and partially miscible systems and showed that a solubility transition, between a drop with surface adsorbed solvent and a drop with solvent mixed into the core occurs before activation in the partially miscible case. 


Deliquescence, where solid particles of soluble salts absorb water directly from the surrounding vapor to form solution droplets, provides an alternative way for particles to grow in the atmosphere. Measurements of micro-sized levitated~\cite{Tang:1994gr,Richardson:1994jz} particles showed deliquescence occurred at a well defined relative humidity corresponding to the vapor pressure where the activity of the water vapor equals the activity of the water in the bulk salt solution, but experiments on nanometer sized particles suggest surface effects can influence the transition for some systems.~\cite{Hameri:2001ew,Biskos:2006hh,Biskos:2006p3170,Hakala:jh} Theoretical studies~\cite{Mirabel:2000p18,Djikaev:2001p758,Djikaev:2002ur,Shchekin:2008p11511,McGraw:2009p10231} show the properties of a surface film that partially dissolves the soluble core, and the nucleation barrier associated with the phase transition, play increasingly important roles in deliquescence as the particles become smaller, suggesting similar effects may become important for small droplets of partially miscible mixtures that can phase separate. 

In this paper, we use a combination of molecular dynamics (MD)  simulations and thermodynamic theory to explore the kinetics and thermodynamics of the condensation of a vapor onto nanoscale, non-volatile liquid drops for systems made up of miscible and partially miscible solvent-solute mixtures. In particular, we are interested in observing features consistent with the presence of a solubility transition in nanoscale droplets. We use a binary mixture of Lennard-Jones particles in our MD simulations because this is a model where the volatility of the component in the drop and the energy of mixing for the two components are easily controlled by adjusting the well depth in the interaction potentials between particles. Measurements of the equilibrium drop size and density of the condensing solvent at the core of the droplet show that the miscible systems alway mix into the core of the drop, while still showing signs of surface enrichment of the volatile component. The partially miscible systems only begin to mix into the droplet core at an onset system volume, once there is a significant amount of vapor condensed onto the drop surface, suggesting the presence of a solubility transition. We also develop a simple capillarity based non-volatile liquid drop model that captures the free energy landscape for a transition between a small and large drop, characteristic of a solubility transition, in the partially miscible system.

The paper is organized as follows: In Section II, we use molecular dynamics simulations to study the dynamics of how the vapor phase condenses onto the drop to form a film for both miscible and partially miscible mixtures. We also measure the equilibrium size and examine the structure of the resulting stable droplet. The non-volatile liquid drop model, based on simple capillarity-type approximations, is developed in Section III to examine the properties of solubility transition observed in partially miscible mixtures. The predictions of the model are compared to the results of our simulations and we also use the model to describe the correct normalization for free energy barriers used in classical nucleation involving nanoscale heterogeneities. Section IV contains our discussion and our conclusions are described in Section V.

\section{Molecular Simulation Studies}
\subsection{Methods}
We use MD simulations in the canonical ensemble to study the condensation of a vapor onto a liquid droplet. The composite system, vapor and droplet, is modelled using a binary mixture of Lennard - Jones particles interacting through the potential,
\begin{equation}
U(r_{ij})=4\epsilon_{ij}\left[\left(\sigma_{ij}/r_{ij}\right)^{12}-\left(\sigma_{ij}/r_{ij}\right)^6\right]{ ,}\\
\label{eq:pot}
\end{equation}
 where $\epsilon_{ij}$ is the energy interaction parameter between species $i$ and $j$, $\sigma_{ij}$ is the particle size interaction between species and the potential is cut, but not shifted, at half the box length. The simulation cell is cubic and we employ periodic boundary conditions. Denoting the volatile solvent and the non-volatile solute as components one and two respectively, we set $\epsilon_{11}=1.0$, $\epsilon_{22}=2.0$ and $\sigma_{11}=\sigma_{22}=\sigma_{12}=1.0$, then vary $\epsilon_{12}$ to control the miscibility of the components.  The energy of mixing parameter,
\begin{equation}
\Lambda^*=(\epsilon_{11}+\epsilon_{22}-2 \epsilon_{12})/\epsilon_{11}\mbox{ ,}\\
\label{eq:lam}
\end{equation}
provides a measure of the energetic drive force for mixing. When $\Lambda^*<0$, particle interactions favor mixing, otherwise they promote phase separation and we study systems with $\Lambda^*=-0.1$ and  $\Lambda^*=0.172$. The ratio $\epsilon_{22}/ \epsilon_{11}$ controls the relative volatility of the components and we have chosen parameters consistent with the DFT model of Talanquer and Oxtoby~\cite{Talanquer:2003p469}  where the supersaturation of the non-volatile phase was $10^5$ times smaller than that of the vapor. The number of volatile particles, $N_1=300$, is maintained for all simulations and we study droplets with $N_2=75,100$ and 150. Our simulations are carried out using the Gromacs package,~\cite{Hess:2008db} where the leap frog integration scheme, with a step size of $\delta t^*=0.002$, is employed to evolve the equations of motion. The velocity rescaling thermostat is used to maintain the system at a reduced temperature, $T^*=kT/\epsilon_{11}=0.8$, where $k$ is the Boltzmann constant. All quantities are reported in reduced units.

\begin{figure}[t]
\includegraphics[width=3.5in]{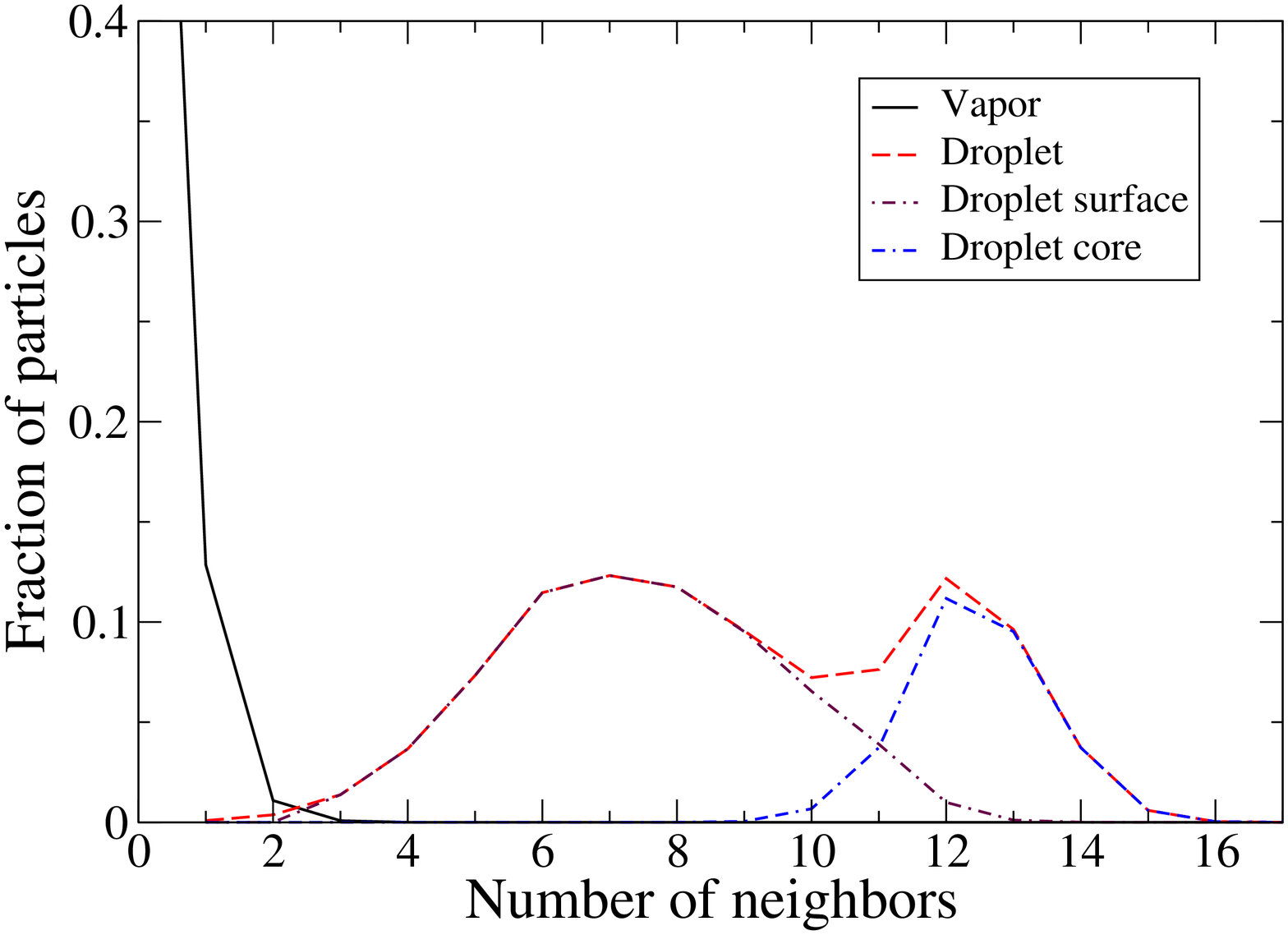}
\caption{Neighbor distribution for particles in the vapor and droplet phase.}
\label{fig:nabs}
\end{figure}

Cluster criteria that allow us to follow the evolution of the droplet as a function of time were developed by measuring the nearest neighbor distributions in both the pure volatile vapor system and in the isolated pure component two droplet phase. Particles of the vapor were initially placed randomly in the simulation cell, with the restriction that no two particles were closer than $1.5\sigma_{11}$, then the system was equilibrated for $10^6$ time steps before configurations were sampled every 10000 time steps for up to $10^7$ time steps. Simulations of the isolated droplet were initialized using a cluster arranged in a compact, body-centered cubic structure that was allowed to equilibrate for $10^7$ time steps. Configurations were then sampled in the same way as the vapor. Fig.~\ref{fig:nabs} shows the fraction of particles with a given number of neighbors within $1.5\sigma_{11}$ for both phases. The distribution of the droplet phase exhibits two distinct peaks that were decomposed into distributions associated with particles at the core of the droplet and those on the surface, using the cone~\cite{Wang:2005ho} method to identify surface atoms. Most core particles have 12 neighbors, which is consistent with the nearest neighbor distribution of the bulk Lennard-Jones fluid with $\epsilon_{ii}=1$~\cite{tenWolde:1998p9475}, but the core distribution measured here is narrower because our non-volatile component has a stronger interaction with $\epsilon_{22}=2$. The distribution for the surface atoms peaks at seven nearest neighbors and marginally overlaps the distribution of the vapor phase, which has no particles with more than three neighbors. On the basis of these results, we identify liquid-like particles as those with three or more neighbors and consider two particles to be part of the same liquid cluster if they are neighbors.

Throughout our simulations we follow three cluster based quantities: i) the size of the droplet, which is taken to be the largest cluster of liquid-like particles in the system and can contain both components, ii) the total number of component one particles in this largest cluster, and iii) the size and number of component one clusters that are part of the largest droplet. At each volume, $V$, studied, we generate the starting configuration by initially equilibrating the isolated component two droplet in the container, then sequentially add the component one particles to the system in random locations, ensuring they are not closer than $1.5\sigma_{11}$ to any other particle. Simulations were then run for $10^7$ time steps. We also measure the equilibrium radial density distribution of each of the components in the droplet from the center of mass of the droplet.

\subsection{Results}
In the absence of the vapor phase, the pure component two droplets with $N_2=100$ and 150 remain stable over the entire volume range studied and we only see the evaporation of five to six  particles from the droplet at the largest volumes studied, confirming that the strong $\epsilon_{22}$ interaction keeps the volatility of the droplet low. The $N_2=75$ droplet begins to show significant evaporation above $V/\sigma^3=3\times 10^5$ and these volumes are not included in our study.

\begin{figure}[t]
\includegraphics[width=3.5in]{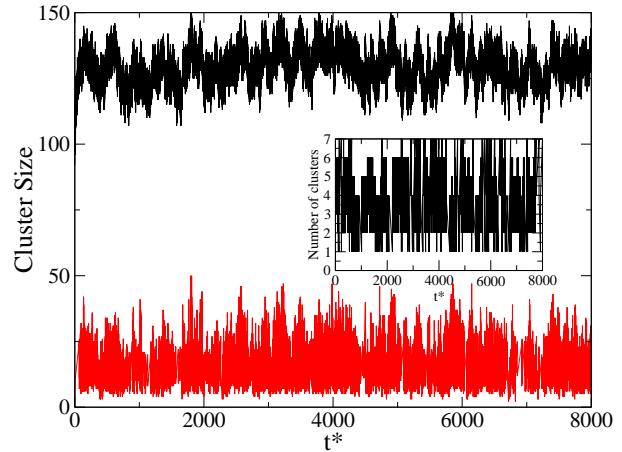}
\caption{Size of the droplet (black line) and size of the largest component one cluster attached to the droplet (red line) for a system with $N_2=100$, $\Lambda^*=0.172$ at $V/\sigma^3=2.5\times 10^5$ as a function of reduced time $t^*$. Inset: Number of clusters of component one attached to drop.}
\label{fig:trajsub}
\end{figure}

\begin{figure}[t]
\includegraphics[width=3.5in]{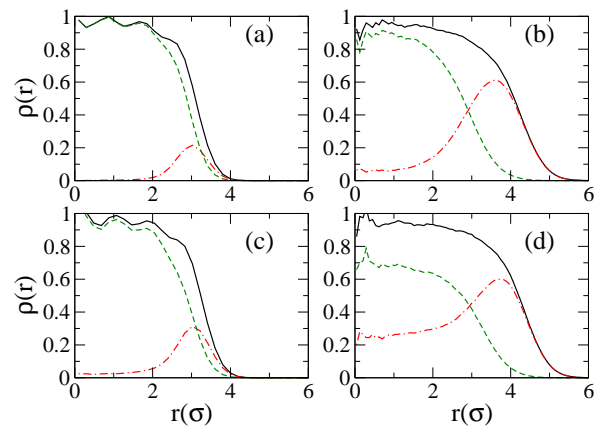}
\caption{{Radial density distributions for the complete droplet (black solid line), component one (red dot-dashed line) and component two (green dashed line) for (a) $\Lambda^*=0.172$, $V/\sigma^3=2.0\times 10^5$, (b) $\Lambda^*=0.172$, $V/\sigma^3=1.5\times 10^4$, (c) $\Lambda^*=-0.1$, $V/\sigma^3=2.0\times 10^5$ and (d) $\Lambda^*=-0.1$, $V/\sigma^3=1.5\times 10^4$.}}
\label{fig:radial}
\end{figure}

In the presence of vapor, the growth of the drop exhibits two distinct time trajectories, depending on the volume of the system. Figure~\ref{fig:trajsub} shows the growth of the droplet, the growth of the largest component one cluster and the number of component one clusters on the drop as a  function of time (Insert).When $V$ is large, the size of the droplet only increases by a small amount and the vapor essentially remains stable. The growth occurs rapidly, then the droplet fluctuates around its equilibrium size, losing and gaining component one particles in a dynamic equilibrium with the vapor phase. The radial density distributions (Figures~\ref{fig:radial}(a) and (c)) show that the component one particles are mainly located in the surface region of the droplet with a small amount of mixing into the core of the droplet when $\Lambda^*=-0.1$. However, there are not enough particles to form a complete monolayer and we see an equilibrium number of component one liquid clusters distributed over the complete cluster surface. This is observed for both values of $\Lambda^*$ studied. Clarke et al~\cite{CLARKE:1994p13243} also observed the submonolayer wetting of droplets in their study of the phase diagram for equimolar binary Lennard-Jones clusters for similar interaction parameters. The fluctuations in the size of the largest component one liquid clusters correlates with the fluctuations in the total size of the droplet, suggesting these clusters grow and shrink by gaining and losing particles to the vapor, although some coalescence between clusters on the droplet surface is expected to be occurring.

\begin{figure}[t]
\includegraphics[width=3.5in]{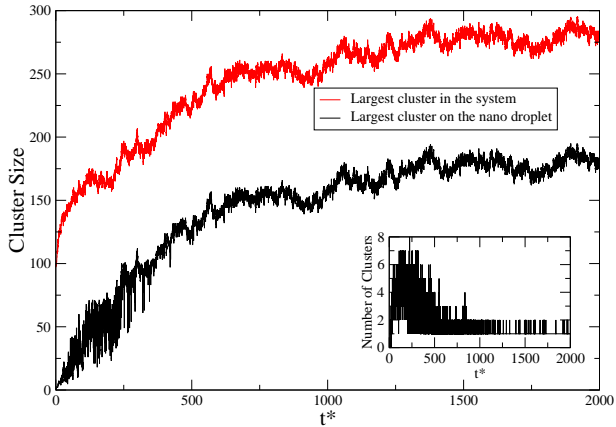}
\caption{Size of the droplet (black line) and size of the largest component one cluster attached to the droplet (red line) for a system with $N_2=100$, $\Lambda^*=0.172$ at $V=1.5\times 10^4$ as a function of reduced time $t^*$. Inset: Number of clusters of component one attached to drop.}
\label{fig:trajmono}
\end{figure}

When $V$ is decreased, the vapor spontaneously condenses onto the droplet, causing it to grow. Figure~\ref{fig:trajmono} shows that the initial condensation onto the droplet leads to a rapid increase in the number and size of component one liquid clusters on the surface of the droplet, but the limited surface area means that the clusters start to interact. In particular, we note that the large fluctuations in the size of the largest component one liquid cluster on the droplet are decoupled from the fluctuations in the total size of the droplet, indicating clusters on the surface are coalescing and breaking up again as the film grows. Eventually, the fluctuations decrease as all the clusters grow and coalesce to form a single cluster representing the completed film. The droplet is unable to grow indefinitely because the total number of component one particles in the simulation container remains fixed in the canonical ensemble and the vapor pressure necessarily decreases until a new equilibrium is established with the enlarged droplet.

\begin{figure}[t]
\includegraphics[width=3.5in]{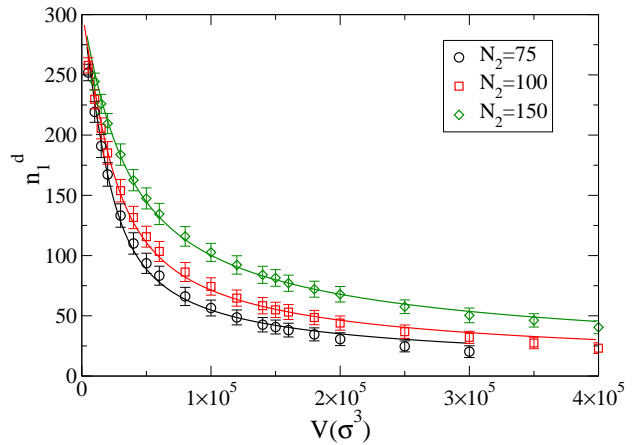}
\caption{$n_1^d$ as a function of $V$ for non-volatile droplets with $\Lambda^*=-0.1$ and sizes $N_2=75,100$ and 150. The error bars represent the standard deviation of $n_1^d$ and the solid lines are the best fits to the data using the non-volatile liquid drop model described in Section III.}
\label{fig:film155}
\end{figure}

\begin{figure}[t]
\includegraphics[width=3.5in]{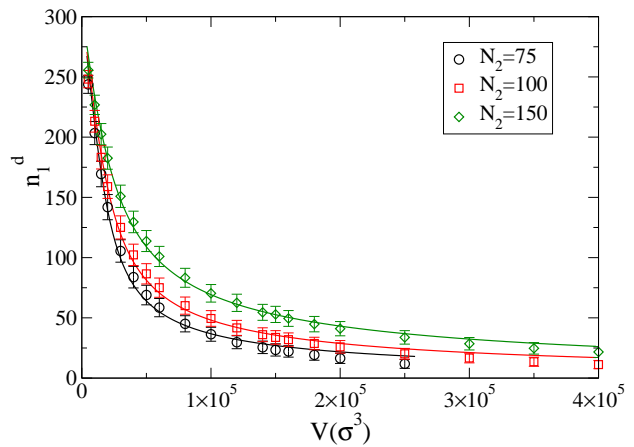}
\caption{$n_1^d$ as a function of $V$ for non-volatile droplets with $\Lambda^*=0.172$ and sizes $N_2=75,100$ and 150. The error bars represent the standard deviation of $n_1^d$ and the solid lines are the best fits to the data using the non-volatile liquid drop model described in Section III.}
\label{fig:film141}
\end{figure}

The value of $\Lambda^*$ has a strong influence on the distribution of the components in the droplet with the radial density distributions (Figure~\ref{fig:radial}) showing that lower values of the mixing parameter lead to greater mixing in the core of the droplet. However, even with  $\Lambda^* < 0$, which represents the point where mixing should be energetically favorable, we see a significant degree of surface enrichment of the volatile solvent at the drop-vapor interface. Figures~\ref{fig:film155} and ~\ref{fig:film141} show the number of component one particles contained in the droplet, $n_1^d$, as a function of the volume of the system. The droplets formed from the miscible mixture generally grow larger than the partially miscible mixture, for drops of the same size, and the larger drops also grow more than the smaller ones, as we might expect. For all systems studied, $n_1^d$ varies continuously over the full range of $V$ studied.

\begin{figure}[t]
\includegraphics[width=3.5in]{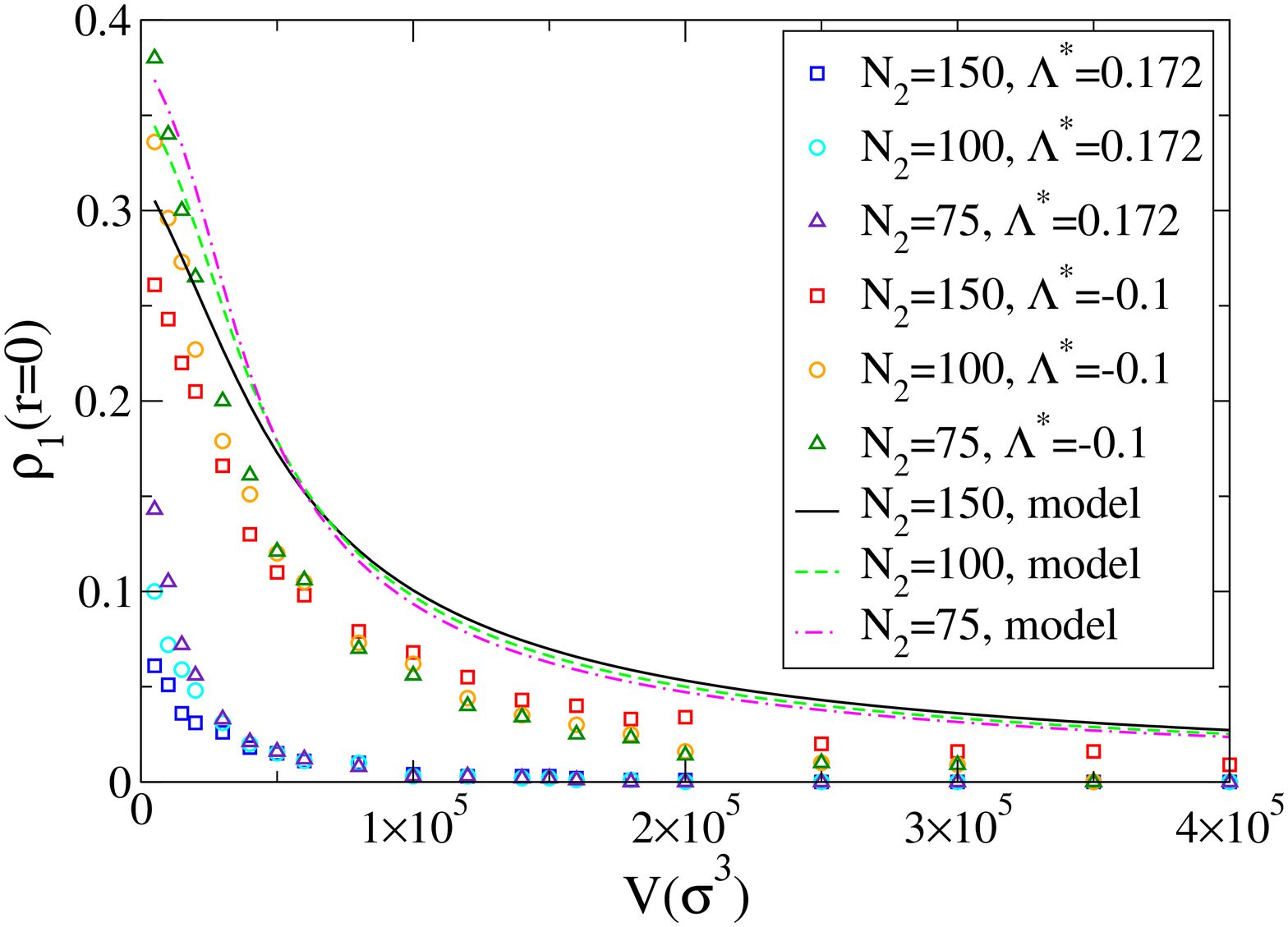}
\caption{Core density $\rho_1(r=0)$ as a function of $V$. The points represent simulation data and the solid lines represent the predictions of the non-volatile liquid drop model described in Section III, with $b_0=0$.}
\label{fig:cd}
\end{figure}

One of the key challenges we face in studying these systems is defining and identifying contributions to $n_1^d$ that arise from adsorption of component one at the non-volatile droplet surface and the mixing of components in the drop. To make this distinction, we measure the extent of mixing into the core of the drop by plotting the density of component one at the core of the drop $\rho_1(r=0)$, obtained from measuring the equilibrium density profiles at $r=0$, as a function of $V$ in Fig.~\ref{fig:cd}. For $\Lambda^*=-0.1$, $\rho_1(r=0)>0$ at all $V$ and varies continuously, indicating the components always mix to some degree. However, we see a change in the $N_2$ dependence of $\rho_1(r=0)$ compared to that observed for $n_1^d$ in Figs.~\ref{fig:film155} and \ref{fig:film141}. Larger non-volatile drops have a higher component one core density when $V$ is large, but the core density of the smaller drops increases faster as $V$ decreases, leading to an inversion of the $\rho_1(r=0)$  dependence on $N_2$, with small drops having greater core densities. The same trend is observed for cases with $\Lambda^*=0.172$ at small system volumes, but at $V/\sigma^3>1.5\times 10^5$, the core density goes to zero, which suggests that the $n_1^d$ particles found in the drop at these volumes can be described as being surface adsorbed. The penetration of component one into the core of the particle, below a specific volume, occurs once more than a monolayer of material is condensed onto the drop and suggests the presence of a solubility transition as a function of volume of the system.

\section{Non-volatile liquid drop model}

\subsection{Model Details}
In this section, we develop the non-volatile liquid drop model, which is an extension of the liquid drop models used to study homogeneous nucleation~\cite{Reguera:2003p5}, binary nucleation~\cite{Reguera:2003p7019} and heterogeneous nucleation.~\cite{Inci:2011hr} Preliminary results for this model were first published in Ref~\cite{Inci:2013bt}. The model consists of a canonical ensemble of $N_1$ particles of the volatile species and $N_2$ particles of the non-volatile species, contained in a fixed volume, $V$, at a fixed temperature, $T$. All $N_2$ particles of the non-volatile species are contained in the spherical droplet phase, along with $n_1^d$ particles of the volatile species. The remaining $n_1^v=N_1-n_1^d$ particles of species one are in the vapor phase which is treated as an ideal gas. At constant $N_1,N_2,V,T$, the Helmholtz free energy, $F$, is the appropriate thermodynamic potential, with variations in $F$ being given by
\begin{equation}
dF=dU-TdS\mbox{,}\\
\label{eq:df0}
\end{equation}
where $U=U^v+U^d$ is the internal energy, $S=S^v+S^d$ is the entropy, and we have denoted quantities relating to the vapor phase and droplet phase with the superscripts $v$ and $d$, respectively. Variations in $U$ are given by
\begin{equation}
dU^v=TdS^v-p^vdV^v+\mu_1^vdn_1^v\mbox{,}\\
\label{eq:duv}
\end{equation}
\begin{equation}
dU^d=TdS^d-p^ddV^d+\mu_1^ddn_1^d+\gamma dA^d\mbox{,}\\
\label{eq:dud}
\end{equation}
where $p^v$ and $p^d$ are the pressures in the respective phases, $\mu_1^v$ is the chemical potential of component one in the vapor phase, $\mu_1^d$ is the chemical potential of component one in the drop phase, $V^d=\nu_1n_1^d+\nu_2 N_2$ is the volume occupied by the drop and $V^v=V-V^d$ is the volume accessible to the vapor. Here, $\nu_i$ is the molecular volume of component $i$ in the bulk liquid phase. We will assume that the drop-vapor interface is sharp, consistent with the capillarity approximation, so that the surface area of the drop is given by $A^d=4\pi R^2$, where $R$ is the radius of the drop, and $\gamma$ is the bulk planar surface tension. It should be noted that Eq.~\ref{eq:dud} does not contain any chemical potential terms for species two because the non-volatile component does not exchange particles with the vapor phase, yielding  $dn_2^d=0$. However, in principle, $dU^d$ should include a term corresponding to the work required to transfer particles from the pure non-volatile drop to the mixed drop. Classical nucleation theory (CNT) generally assumes this term is large and negative but independent of the radius of the drop so that it is ignored as it does not affect the nucleation rate~\cite{Sear:2008p5489}. As our derivation proceeds, we will see that neglecting this term in Eq.~\ref{eq:dud} amounts to assuming that component two behaves ideally in solution, even if we have included non-ideal behavior for component one in the drop phase.

Using the conservation conditions $dV^v=-dV^d$ and $dn_1^v=-dn_1^d$ , along with the relation $dA^d=2dV^d/R$, in Eqs.~\ref{eq:df0}-\ref{eq:dud} yields
\begin{equation}
dF=\left(p^v-p^d+\frac{2\gamma}{R}\right)dV^d+\left(\mu_1^d-\mu_1^v\right) dn_1^d\mbox{ ,}\\
\label{eq:df1}
\end{equation}
which gives the equilibrium conditions satisfying $dF=0$ as $\mu_1^d=\mu_1^v$ and  $p^d-p^v=2\gamma/R$. To obtain a more detailed model, we need to obtain relations describing the chemical potential terms. Integrating the Gibbs-Duhem relation for the vapor phase, $S^vdT-V^vdp^v+n_1^vd\mu_1^v=0$, at constant $T$ gives
\begin{equation}
\mu_1^v(p^v)-\mu_1^{eq}(p_1^{eq})=kT\ln\frac{p^v}{p_1^{eq}}\mbox{,}\\
\label{eq:muideal}
\end{equation}
where we have used the coexistence pressure of the vapor in contact with the pure fluid via a planar interface, $p_1^{eq}$, as the reference state. The Gibbs-Duhem relation for the drop and its associated surface is $S_1^ddT-V^d dp^d+n_1^d d\mu_1^d+A^d d\gamma=0$. Assuming that the drop is incompressible, and that the surface tension is independent of both the pressure and the composition of the drop,  yields
\begin{equation}
\mu_1^d(p^d,T,x_1)-\mu_1^d(p_1^{eq},T,x_1)=\nu_1(p^d-p_1^{eq})\mbox{,}\\
\label{eq:mu1a}
\end{equation}
where $x_1=n_1^d/(n_1^d+N_2)$ is the mole fraction of component one in the drop.

One way to capture the non-ideal nature of the system studied in our simulations is to treat component one in the drop as a regular solution,~\cite{Hill:1987fk} which is based on the Bragg-Williams lattice approximation. This assumes that the drop mixes uniformly so the entropy of mixing is the same as the ideal solution, but the enthalpy of mixing is dependent on $x_1$. The chemical potential of component one at $p_1^{eq},T$ and $x_1$ is then described by
\begin{equation}
\mu_1^d(p_1^{eq},T,x_1)=\mu^0+kT\ln x_1 + b_0(1-x_1)^2\mbox{,}\\
\label{eq:mureal}
\end{equation}
where $\mu^0=\mu_1^d(p_1^{eq},T,x_1=1)$ is the chemical potential of pure component one liquid at $p_1^{eq}$ and $b_0$ accounts for the interaction between components. When $b_0$ is set to zero, Eq.~\ref{eq:mureal} reduces to the expression for an ideal solution, while positive and negative values correspond to repulsive and attractive interactions respectively.

Using Eqs.~\ref{eq:muideal}-\ref{eq:mureal} in Eq.~\ref{eq:df1} and noting $dV^d=\nu_1 dn_1^d$ yields
\begin{equation}
dF=\left[\nu_1(p^v-p_1^{eq})-kT\ln\frac{p^v}{x_1 p_1^{eq}}+\frac{2\nu_1\gamma}{R}+b_0(1-x_1)^2\right]dn_1^d\mbox{.}\\
\label{eq:dfdn}
\end{equation}
Equating the term inside the brackets of Eq.~\ref{eq:dfdn} to zero then gives us the Kelvin relation for the binary drop,
\begin{equation}
\frac{p^v}{p_1^{eq}}=a_1\exp \left[\frac{2\nu_1\gamma}{kTR}\right]\mbox{,}\\
\label{eq:kelvin}
\end{equation}
where $a_1=x_1\exp[b_0(1-x_1)^2/kT]$ is the activity of component one in the drop and the first term has been ignored because it is generally small. Reiss and Koper\cite{Reiss:1995p7018} also obtained Eq.~\ref{eq:kelvin} using a different approach in an open system. Finally, Eq.~\ref{eq:dfdn} is integrated with respect to $n_1^d$ to obtain the free energy of forming the drop,
\begin{equation}
\begin{array}{lll}
\Delta F&=&F(n_1^d)-F(0)\nonumber\\
&=& n_1^d\left[ kT-\nu_1^d p_1^{eq}+b_0(1-x_1)\right]+N_2 kT\ln (1-x_1)\nonumber\\
&+&N_1 kT\ln \frac{p^v}{p_1^0}-n_1^d kT\ln\frac{p_1^v}{x_1 P_1^{eq}}+\gamma[A^d(n_1^d)-A^d(0)]\nonumber
\mbox{,}
\label{eq:fefinal}
\end{array}
\end{equation}
where $p_1^0=N_1kT/(V-V^d)$ is the pressure of the vapor before any particles have condensed onto the drop and the second term of the right hand side of the equation is the entropy of mixing for component two.

\subsection{Results}
\begin{figure}[t]
\includegraphics[width=3.5in]{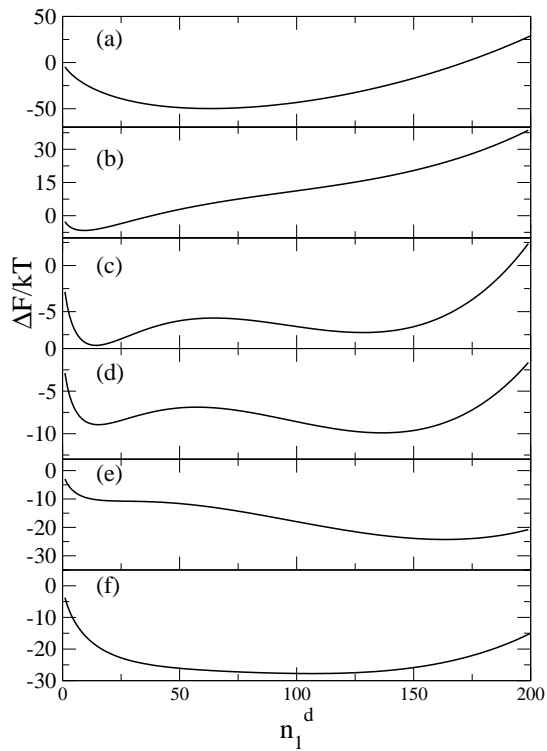}
\caption{Free energy landscape as a function of $n_1^d$ for a droplet with (a) $b_0=0$, $N_2=100$, $V/\sigma^3=30000$, then for droplets with $b_0=3$ and $N_2=100$ at (b) $V/\sigma^3=12000$, (c) $V//\sigma^3=10000$, (d) $V/\sigma^3=9800$, (e) $V/\sigma^3=8900$, and (f) $N_2=200$, $V/\sigma^3=10000$.}
\label{fig:feld}
\end{figure}

We begin this section by exploring the general features of the free energy surface described by Eq.~\ref{eq:fefinal} as a function $V$ and $n_1^d$, using the thermodynamic parameters for argon~\cite{Baidakov:2007p11419} where appropriate and assuming $\nu_1=\nu_2$. We also set $N_1=300$ to be consistent with our simulations. When $b_0$ is small or negative, the entropy of mixing for the two components dominates the free energy and we find there is a single free energy minimum, corresponding to the spontaneous formation of a mixed drop, for all system volumes (Figure~\ref{fig:feld}(a)). As $V$ decreases, the minimum simply moves to larger $n_1^d$ as more vapor condenses and the drop grows.

Figures~\ref{fig:feld}(b)-(e) show that the evolution of the free energy surface as a function of volume, for a system with  larger $b_0$, traces out a hysteresis loop similar to the one observed in the deliquescence and efflorescence of soluble salt particles. At large $V$, we see the spontaneous absorption of a few component one particles to form a small mixed drop. As $V$ is decreased, a minimum appears at larger $n_1^d$, corresponding to a drop that has absorbed a significant fraction of the volatile solvent component, dissolving the non-volatile solute. Initially the minimum is metastable relative to the small drop, but it eventually becomes the most stable state.  The free energy minima for the two drops is separated by a free energy maximum that is an unstable equilibrium solution to the Kelvin equation and the size of the drop at the maximum is the critical nucleus for the solubility transition. At smaller $V$, we reach the limit of stability of the non-volatile drop, which then dissolves spontaneously. This point represents the activated drop described in K\"{o}hler theory.

As the initial size of the non-volatile drop decreases, the minimum in the free energy for the small drop moves to smaller $n_1^d$ and becomes shallower. In the limit of $N_2\rightarrow 0$ Eq.~\ref{eq:fefinal}, reduces to the free energy expression for the original homogeneous liquid drop model~\cite{Reguera:2003p5}. For large $N_2$, the free energy surface is once again characterized by a single broad minimum (Figure~\ref{fig:feld}(f)) as the entropy of mixing again becomes the most dominant term in the free energy.

\begin{figure}[t]
\includegraphics[width=3.5in]{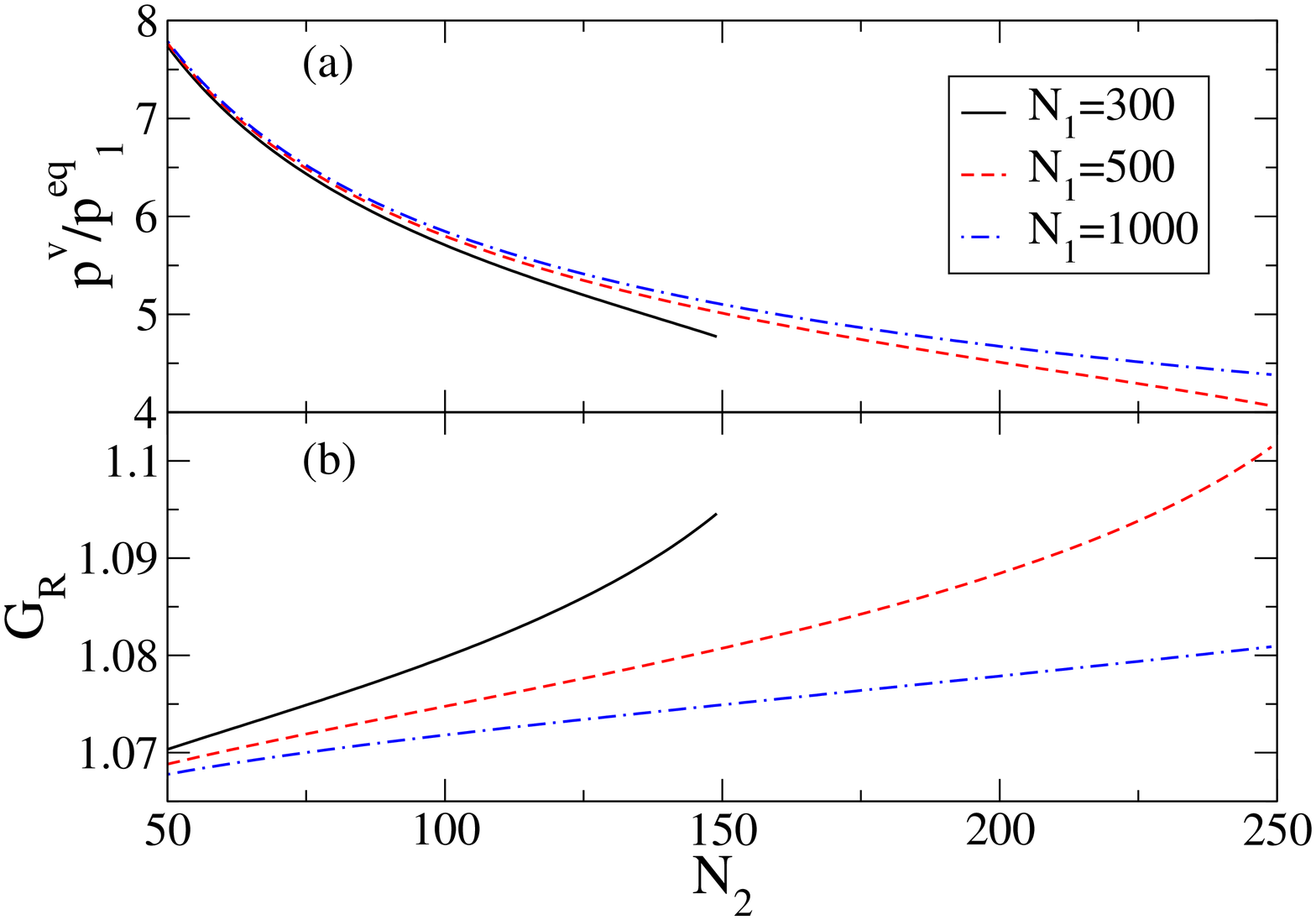}
\caption{(a) The vapor pressure for the drop, given by Eq~\ref{eq:kelvin}, at the activation point as a function of $N_2$ for different number of initial values of $N_1$. (b) Growth factor, $G_R$, of the drop at the activation point as a function of $N_2$.}
\label{fig:stability}
\end{figure}

To study the effect of the size of the non-volatile droplet on the location of the transition between small and large drops, we assume, as in the case of K\"{o}hler activation theory, that the transition occurs at the point the small drop becomes unstable and satisfies the conditions $\partial F/\partial n_1^d=0$, and $\partial^2 F/\partial n_1^{d2}=0$. 
Figure~\ref{fig:stability}(a) shows that the pressure of the vapor surrounding the unstable drop, relative to the bulk equilibrium vapor pressure, as given by Eq.~\ref{eq:kelvin}, increases with the decreasing $N_2$. We also see the that the growth factor of the drop, $G_R=R/R_0$ where $R_0$ is the size of the pure non-volatile drop, decreases with decreasing $N_2$ (Figure~\ref{fig:stability}(b)). The results of the model are mildly dependent on $N_1$ because the vapor phase becomes depleted as the droplet grows, but these effects decrease as the system size increases.

Figures~\ref{fig:film155} and \ref{fig:film141} show fits of the liquid drop model to our simulation data, where we have used $b_0$ and $\gamma$ as fit parameters, while fixing the remaining parameters. The model fits the data well for both values of $\Lambda^*$, and all non-volatile drop sizes, over the entire range of volumes studied. The values of $\gamma^*$ range from 1.59 -- 1.80 and 1.92 -- 2.36 for $\Lambda^*=-0.1$ and $0.174$ respectively. These values are higher than the surface tension for the pure fluid of component one and are physically reasonable, but we would expect the true surface tension of the drop to be dependent on the mole fraction of the components and this effect has been ignored in our model. We also find that the fit values for $b_0$ are negative. This reflects the fact that the model assumes all $n_1^d$ condensed particles are uniformly distributed in the drop rather than having some partitioned to the surface, so it over--estimates the degree to which the particles like to mix. A key feature of the models is that it predicts the transition between small and large drops should be accompanied by a discontinuous increase in $n_1^d$ as a function of $V$. The simulation trajectories show that the thin films form spontaneously and there is no clear sign of nucleation like behavior or of the expected discontinuity in the equilibrium droplet size, but these may be obscured to some degree by surface absorption.

We also compare the model predictions for the core density using $\rho_1(r=0)=\rho_1=n_1^d/(\nu_1n_1^d+\nu_2N_2)$ because it is assumed the components are uniformly mixed. Figure~\ref{fig:cd} shows that the model, assuming ideal mixing ($b_0=0$) correctly predicts the dependency of $\rho_1(r=0)$ on $N_2$, including the inversion of the trend as a function of $V$. Fitting the model to the data using $b_0$ and $\gamma$ as fit parameters yield excellent looking curves, but the two fit parameters become highly anti-correlated and particle size dependent. These fits have not been included here as the values of the parameters they yield appear unphysical.

\subsection{Nucleation}
K\"{o}hler activation theory, as used in the last section, successfully describes the location of the transition in large drops but nucleation becomes an increasingly important mechanism for droplet growth as particle size decreases~\cite{Djikaev:2001p758,Djikaev:2002ur,McGraw:2009p10231} and we would expect the transition to occur at lower vapor pressures because droplets can get over the barrier in an activated process before the limit of stability is reached. In this section, we will use the non-volatile liquid drop model to examine the main features of nucleation in systems that exhibit a free energy minimum along the reaction coordinate and the effect of particle size on the nucleation barrier.

\begin{figure}[t]
\includegraphics[width=3.5in]{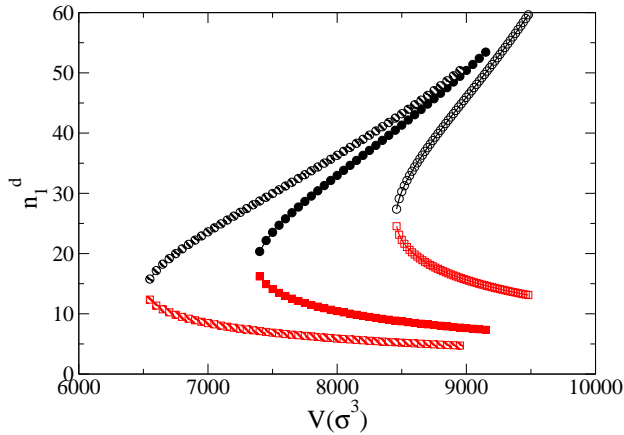}
\caption{The number of solvent atoms in the droplet at the free energy minimum, $n_{min}$ (squares), and at the free energy maximum, $n^*$ (circles), for droplets with $N_2=100$ (open symbols), $N_2=75$ (filled symbols) and $N_2=50$ (striped symbols).}
\label{fig:nminnmax}
\end{figure}

The nucleation process of interest is the dissolution of the small drop, which occurs when the drop absorbs enough of the solvent phase to grow larger than the critical drop size, $n^*$, located at the maximum in the free energy curve. Figure~\ref{fig:nminnmax} plots the size of the critical droplet and the size of the small droplet at the minimum over the range of volumes where the small drop is metastable. As the system volume decreases, $n^*$ decreases and $n_{min}$ increases until they converge at the limit of stability, which represents the activation point. These plots show that the range of $V$ over which the small mixed drop is metastable increases for decreasing $N_2$, which helps explain why nucleation becomes more important as the non-volatile particle gets smaller. When $N_2$ is large, a small change in $V$ will rapidly move the system beyond the limit of stability, while the smaller particles require much larger changes in $V$.

Classical nucleation theory (CNT), gives the rate of drops going over the barrier as
\begin{equation}
J=A\exp(-\Delta F^*/kT)\mbox{,}\\
\label{eq:ratecnt}
\end{equation}
where $\Delta F^*/kT$ is the height of the free energy barrier and $A$ is a pre-exponential factor containing information about the dynamics. When the free energy surface contains a local minimum, the free energy barrier used in Eq.~\ref{eq:ratecnt} is usually defined as the difference in free energy between the maximum and the minimum,~\cite{Talanquer:2003p469,Sear:2008p5489} $\Delta F_{mm}=\Delta F(n^*)-\Delta F(n_{min})$. However, Scheifele et al.~\cite{Scheifele:2013uo} showed that $\Delta F_{mm}/kT$ did not correctly predict the rate for the heterogeneous nucleation of the two dimensional Ising model onto a nanoscale impurity and highlighted the fact that the exponential term in the rate expression is really a surrogate for the probability of finding the metastable  drop at the transition state, $P(n^*)$.  Our model exhibits a free energy minimum and should be extensive in the number of non-volatile particles, so we would anticipate that the same analysis is needed here to correctly predict the rate.

The free energy that provides the probability of finding a drop containing $n_1^d$ particles can be expressed
\begin{equation}
\Delta F_0(n_1^d)/kT=-\ln P(n_1^d)=-\ln\frac{Q(n_1^d)}{Q(\mbox{met})}\mbox{,}\\
\label{eq:fnorm}
\end{equation}
where $Q(n_1^d)$ is the partition function of the drop with $n_1^d$ and $Q(\mbox{met})$ is the partition function of the metastable system,
\begin{equation}
Q(\mbox{met})=\sum_{n_1^d=0}^{n_1^d=n^*}Q(n_1^d)\mbox{.}\\
\label{eq:qmet}
\end{equation}
The free energy that should appear in the rate expression is then given by $\Delta F_0(n^*)/kT$ and represents the work required to constrain the metastable droplet to its critical size. In the context of the thermodynamic, capillarity based model developed here, $\Delta F_0(n_1^d)/kT$ can be calculated by renormalizing the free energy given by Eq.~\ref{eq:fefinal} so that,
\begin{equation}
P(n_1^d)=\frac{\exp(-\Delta F(n_1^d)/kT)}{\sum_{n_1^d=0}^{n_1^d=n^*}\exp(-\Delta F(n_1^d)/kT)}\mbox{,}\\
\label{eq:pnorm}
\end{equation}
which ensures $\sum_{n_1^d=0}^{n_1^d=n^*}P(n_1^d)=1$. $\Delta F_0(n_1^d)/kT$ can then be obtained from the left hand equality in Eq.~\ref{eq:fnorm}. Implicit in Eq.~\ref{eq:pnorm} is the assumption that the capillarity model, which gives rise to $\Delta F(n_1^d)$, describes all the microscopic states of the partition function, $Q(n_1^d)$, and that $n_1^d$ serves as an order parameter that can be used to rigorously sum over all the possible states. 

\begin{figure}[t]
\includegraphics[width=3.5in]{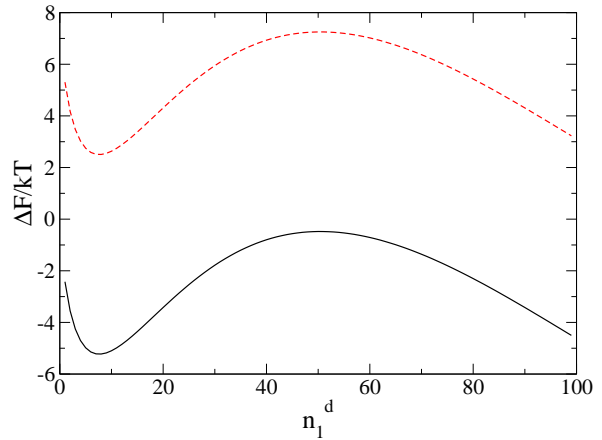}
\centering
\caption{The free energy before (solid line) and after (dashed line) renormalization for nucleation.}
\label{fig:dfcomp}
\end{figure}

\begin{figure}[t]
\includegraphics[width=3.5in]{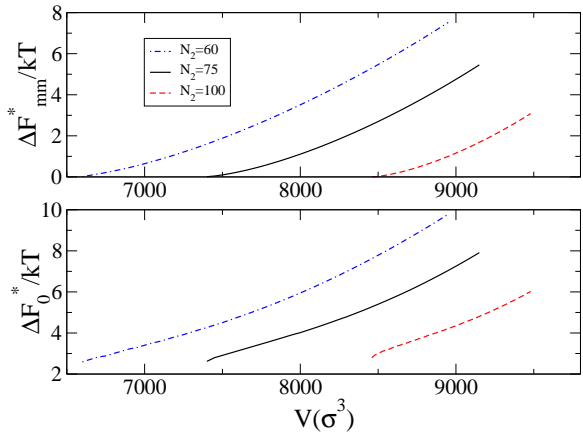}
\centering
\caption{Nucleation barriers calculated using (top) the difference in free energy from maximum to minimum, $\Delta F^*_{mm}/kT$, and (bottom) the renormalized free energy $\Delta F^*_{0}/kT$, as a function of $V$ for systems with $N_2=60, 75$ and 100.}
\label{fig:nucbar}
\end{figure}

Figure~\ref{fig:dfcomp} shows that the effect of the renormalization is to shift the free energy curve vertically, without changing its shape, because the demoninator in Eq.~\ref{eq:pnorm} results in a constant term that is applied to the free energy of all the states. $\Delta F_0(n_1^d)/kT)$ is positive for all the drop sizes in the metastable region since it always takes work to constrain the system to the subset of states, but the values for $n_1^d > n^*$ have no real meaning because they are connected to the stable state and have not been included in the metastable drop partition function. A comparison of the nucleation free energy barriers predicted by $\Delta F_{mm}$ and Eqs.~\ref{eq:fnorm} -- \ref{eq:pnorm}  (see Figure~\ref{fig:nucbar}) shows that $\Delta F_0^*/kT$ is generally $2-3\mbox{ } kT$ higher than the minimum to maximum barrier which can lead to orders of magnitudes of difference in the nucleation rates predicted by the two definitions. 

$\Delta F_0^*/kT$ also remains finite at the limit of stability. This seems counter intuitive because we associate a positive free energy barrier with an activated process, but at the limit of stability the free energy landscape is monotonically decreasing from $n_1^d=0$ so the drop grows spontaneously. This remains true, even after renormalization. However, in reality, the metastable region ceases to be defined at the limit of stability and the barrier is only defined as the limit of stability is approached from above. The finite barrier is then a result of the fact that $n^*\rightarrow n_n$ at a finite value of $n_1^d$, because of the minimum in the free energy.

Despite this unusual property, Scheifele et. al.~\cite{Scheifele:2013uo} showed that the free energy given by Eq~\ref{eq:fnorm} correctly predicts the rate of nucleation right down to the limit of stability for their heterogeneous nucleation case. We do not have any independent rate data to test in our current model but our analysis provides a useful example showing how the renormalization can be implemented for a capillarity--based model.

\section{Discussion}
In this work, we have performed a series of molecular dynamics simulations to study the condensation of a vapor onto a non-volatile drop for both miscible and partially miscible binary Lennard-Jones mixtures. In the canonical ensemble, the drop grows spontaneously as the vapor condenses, but eventually it comes to equilibrium as the vapor phase is depleted. When $V$ is large, a submonolayer amount of the vapor is adsorbed onto the drop with the particles being distributed over the droplet surface in small clusters. Some mixing into the core does occur for the miscible mixtures but no mixing is observed for the  partially miscible systems. When $V$ is small enough to cause a substantial number of vapor particles to condense, we see a film growth mechanism that is dominated by cluster-cluster coalescence due to the restricted surface area available on the nanoscale sized drop. This is likely to be a common feature of nucleation and growth mechanisms in nanoscale systems. It is in contrast to the usual mechanisms observed during film formation on macroscopic surfaces that usually occur through the addition and loss of individual particles to and from isolated clusters. Once a monolayer is formed, we also begin to see mixing into the core for the droplet of the partially miscible system, which is a sign that the droplet core has started to dissolve.

We also developed the non-volatile liquid drop model, combined with elements of the regular solution theory, to described the general features of the free energy surface associated with droplet growth in nanoscale systems. The free energy landscape of the model for partially miscible components exhibits a hysteresis loop similar to that observed in deliquescence and efflorescence of small soluble salt particles, caused by the presence of a nucleation barrier between the small drop and large dissolved drop phases. This transition resembles elements of the solubility transition described by Talanquer and Oxtoby~\cite{Talanquer:2003p469} using DFT. However, the DFT model directly includes effects due to surface adsorption where our model ignores this feature, even though our simulations show that these are important.

A number of capillarity based models have been used to study the effects of surface films on the deliquescence of small particles. For example, introducing the disjoining pressure~\cite{Djikaev:2001p758,Shchekin:2008p11511} to account for the interaction between vapor--film and film--solid interfaces is able to stabilize the solution film which would otherwise spontaneously dissolve the salt particle. A thin layer criterion was developed by McGraw and Lewis~\cite{McGraw:2009p10231} that requires the equality of the chemical potentials between the solvent component in the vapor and the film, as well as the equality of the chemical potentials between the salt in the film and the solid particle. Our simulations clearly highlight the importance of surface films in the cases of miscible and partially miscible systems and it may be useful to develop these approaches in the context of the solubility transition in small systems. Finally, experiments~\cite{SEMENIUK:2007jy,Bertram:2011eu,You:ue} have shown that large atmospheric aerosols made from complex mixtures of soluble salts, organics and water lead to increasingly complicated cycles of structural transformations where solubility-like liquid-liquid transitions, involving phase separation of organics from inorganic salt solutions, and the deliquescence of the resulting salt solution all occur as a function of the relative humidity. These studies have focused on large particles, but  it is likely surface effects will further complicate the nature of these transformations as the particles become smaller.

\section{Conclusion}
Understanding nanoscale particle size effects on the dynamics, thermodynamics and physical structure of small atmospheric aerosol particles remains an important challenge. We have shown that molecular dynamics simulations of a simple binary mixture of Lennard-Jones particles are able to capture the key elements of the dynamics and thermodynamics of the condensation of a solvent vapor onto a non-volatile solute particle. In particular, we have shown that cluster-cluster coalescence plays an important role in film formation in nanoscale surfaces and that partially miscible droplets exhibit a solubility transition. We have also shown that a simple capillarity based model also captures the main features of the solubility transition, but more complex models are needed to account for surface adsorption.

\acknowledgements{We would like to thank NSERC for financial support. Computing resources were provided by CFI and WestGrid.}

%

\end{document}